\documentclass[twocolumn,prx,aps,longbibliography]{revtex4-1} 

\usepackage{color} 
\usepackage{natbib}
\usepackage{amsmath} 
\usepackage{multirow}  
\usepackage{graphicx} 
\usepackage{soul}
\usepackage{url}
\usepackage{textcase}
\usepackage{bm}
\usepackage{epstopdf}

\newcommand{\eref}[1]{(\ref{#1})} 
\newcommand{\from}{\leftarrow}

\newcommand{\set}[1]{\{#1\}}

\newcommand{\eps}{\epsilon} 
\newcommand{\rats}{\vec{\pi}} 
 
\newcommand{\ex}[1]{\mathrm{exp}(#1)} 
\newcommand{\defn}[1]{\textbf{#1}} 

\renewcommand{\P}{\mathbf{P}} 
\newcommand{\sij}{\sigma_{ij}} 
\newcommand{\sji}{\sigma_{ji}}

\newcommand{\ra}{\pi} 

\newlength{\figurewidth} 
\ifdim\columnwidth<20cm 
\else 
\fi 
\begin{document} 

\title{Ranking Competitors Using Degree-Neutralized Random Walks}
\author{Seungkyu Shin} 
\affiliation{Graduate School of Culture Technology, Korea Advanced Institute of Science and Technology, Daejeon, Republic of Korea} 
\author{Sebastian E. Ahnert} 
\affiliation{Theory of Condensed Matter, Cavendish Laboratory, CB3 0HE Cambridge, UK} 
\author{Juyong Park} 
\affiliation{Graduate School of Culture Technology, Korea Advanced Institute of Science and Technology, Daejeon, Republic of Korea} 

\begin{abstract}
Competition is ubiquitous in many complex biological, social, and technological systems, playing an integral role in the evolutionary dynamics of the systems. It is often useful to determine the dominance hierarchy or the rankings of the components of the system that compete for survival and success based on the outcomes of the competitions between them. Here we propose a ranking method based on the random walk on the network representing the competitors as nodes and competitions as directed edges with asymmetric weights. We use the edge weights and node degrees to define the gradient on each edge that guides the random walker towards the weaker (or the stronger) node, which enables us to interpret the steady-state occupancy as the measure of the node's weakness (or strength) that is free of unwarranted degree-induced bias. We apply our method to two real-world competition networks and explore the issues of ranking stabilization and prediction accuracy, finding that our method outperforms other methods including the baseline win--loss differential method in sparse networks.
\end{abstract} 

\maketitle 

\section{Introduction} 
Competition is one of the most essential mechanisms for the survival and evolution of species or components in a complex system, be it from the biological, the social, or the technological realm~\cite{darwin2009origin,stuart1993origins,drossel2001biological}.  Therefore in many complex systems a robust and effective ``rating'' or ``ranking'' method for determining the most successful or superior component can be essential for understanding its dynamics~\cite{williams2000simple,park2005network,motegi2012network,park2015bayesian,balinski2010majority}.  It can also contribute to a system's success and confidence: In an enterprise, for instance, a fair competition-and-reward mechanism would be the basis for earning the confidence of its employees and success.  The same argument would apply to an economic or financial institution; the confidence in the fairness of the ratings system by the participants such as investors and customers is crucial for its sustainability and development.

Here we propose a ranking method where the competing species are represented as nodes of a competition network. The most familiar example of a competition network can be found in sports where the nodes represent competing teams, and the edges the competitions or the games played between them~\cite{park2005network}. The ranking of the competitors (teams) in a sport is an issue of much interest, as it functions as the basis of many events (e.g., playoffs) and decisions (e.g., marketing) that could determine its popularity and success.  While the ranking methods differ from sport to sport, they are almost always some type of a generalization of the simplest scheme of counting the wins and losses that can often be insufficient for the purpose of producing satisfactory and useful rankings~\cite{stefani1997survey}.

\begin{figure} 
\includegraphics[width=50mm]{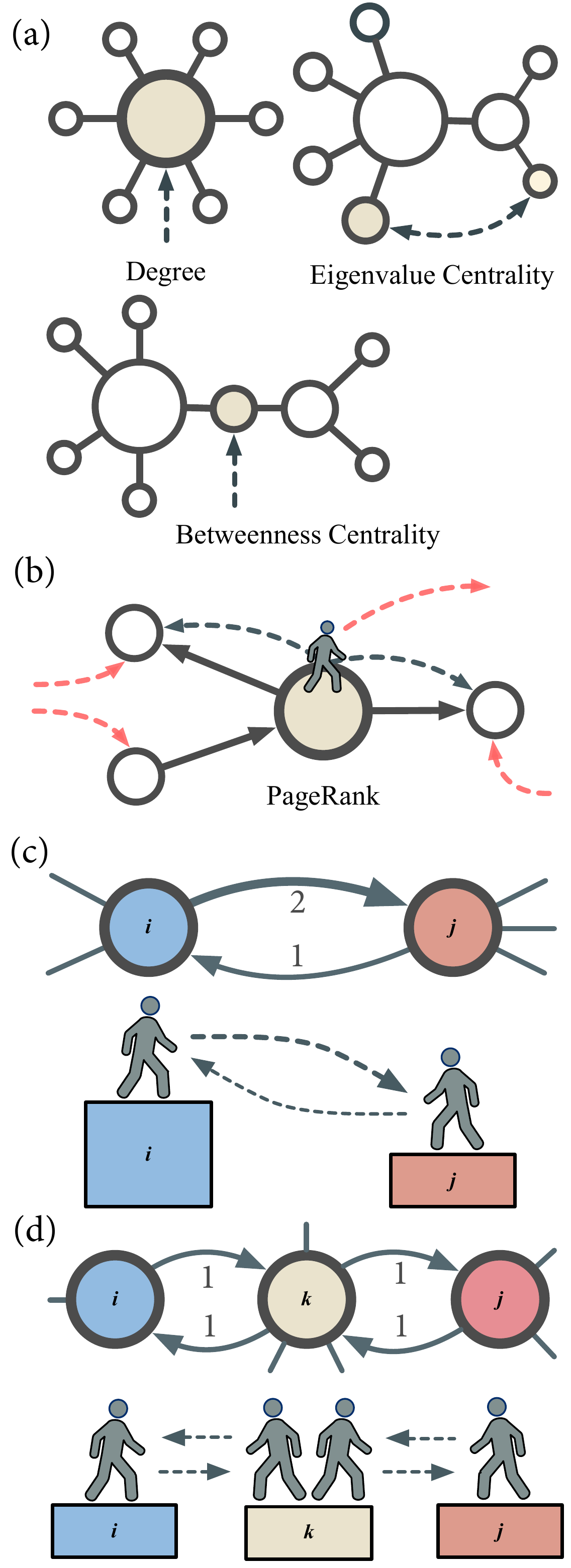} 
\caption{Basic concept of network centralities and our ranking method. (a) Three basic network centralities. The \defn{Degree} is the number of node's neighbors; the shaded node in the most central. The \defn{Eigenvalue Centrality} considers the quality of a connection, so that being connected to a central node raises one's centrality in turn; the larger shaded node is more central than the smaller shaded node, although their degrees are equal. The \defn{Betweenness} quantifies the node's role in acting as an intermediary between nodes by measuring how often it sits on the geodesic (shortest) paths between two nodes; the shaded node, even though its degree is low, is the most central. (b) In PageRank, with probability $\alpha=0.85$ a random walker follows a randomly chosen outgoing link (solid lines) to travel to another node, and with probability $1-\alpha=0.15$ makes a random jump to any node in the network (red dotted lines). The nodes are ranked by their stationary occupation probability.  (c) A competition network is a directed network with weighted directional edges, where the weights can represent the number of wins or the points scored by one node against another. Our ranking method is based on random walk where the edge weights define a gradient between nodes. We use the stationary occupation probability as the measure of node's strength or weakness, depending on the defined directionality of the gradient. (d) A high degree can unfairly favor and penalize a node, necessitating a degree-neutralizing procedure.} 
\label{figure01} 
\end{figure}

The ranking of nodes in a network has a long and rich history of development~\cite{freeman2004development,Katz53a,freeman1979centrality,Newman:2010fk}.  It is most often formulated in the network context as a problem of ``centrality,'' i.e. the measure of a node's prominence or importance deduced from the network structure.  Of many popular centralities (Google's PageRank~\cite{brin1998anatomy} is perhaps the best-known modern example, to be discussed below) in existence, in Fig.~\ref{figure01}~(a) we show three fundamental ones that often serve as bases for more elaborate ones~\cite{wasserman1994social,Newman:2010fk}. The first is the \defn{degree centrality}, or simply the degree, which is the number of nodes connected to the node, rendering the node in the middle the most central.  The second is the \defn{eigenvector centrality}, which is the leading positive eigenvector of the adjacency matrix.  It generalizes the degree taking into account the ``quality'' of a connection, thereby differentiating the two shaded nodes with the same degree $(1)$ in the picture. The third is the \defn{betweenness} or \defn{Freeman centrality} which measures how often a node sits on the shortest path(s) between two nodes~\cite{freeman1977set}. Thus shaded node in the center is the most central using betweenness, although its degree is smaller than those of its neighbors. 

Fig.~\ref{figure01}~(b) shows how PageRank works. Devised for ranking webpages in the Worldwide Web with pages as nodes and hyperlinks as directed edges of a network, it employs the concept of random walk where at each time step the walker moves from a node to another by following a randomly chosen outgoing link with probability $\alpha=0.85$ (called the ``Google alpha''), or jumps to a randomly chosen node (regardless of connection) with probability $1-\alpha=0.15$. Interpreting an incoming edge (i.e. being cited by a webpage) as indicating a node's significance, PageRank is then given by the stationary occupation probabilities for each node under the random walk. The idea of the random walk is used in our proposed method explained below, along with the difference between the two methods.

\section{Methods}
A competition network can be represented as a weighted directed network, as shown in Fig.~\ref{figure01}~(c). We call $s_{ij}$ the weight of the edge to $i$ from $j$, which can be the points scored by $j$ against $i$ in a sports match, the number of times that $j$ beat $i$ in a series of encounters, etc.~\cite{williams2000simple,park2005network} (This is a matter of convention. In ecology, for instance, it is more common to allow an edge point from the prey (loser) to its predator (winner)). Therefore Fig.~\ref{figure01}~(c) might represent the result of soccer game in which team $i$ (left) beat team $j$ by the score of 2:1 (i.e. $s_{ji}=2$ and $s_{ij}=1$).

To determine the global ranking of nodes from the strongest to the weakest based on the weights $\set{s_{ij}}$, we picture a random walker who travels indefinitely from node to node along the network edges that have a slope (gradient) defined by the weights.  Let us, for the time being, assume a downward slope from the winner to the loser; then this will cause the walker to visit the weaker nodes more often, so that we can rank the nodes in the order of the increasing occupation probability $\vec{l}=\set{l_1,l_2,\ldots,l_n}$ where $n$ is the number of nodes in the network, and $\sum_il_i=1$.  We allow, however, the walker to travel up the slope as well as down it, only not as easily.  There are two reasons for this.  First, since the outcome of a real competition event is inherently stochastic, it may well be the case that a truly weaker node may have defeated a stronger opponent, which we call an ``upset'' in sports parlance. Second, such a bidirectional travel prevents the pathological cases where the random walker gets stuck at a node with no exits (i.e. a node that has lost all contests).

A possible form for the transition probability $t_{ij}\equiv t_{i\from j}$ between connected nodes (i.e. the adjacency matrix element $\sigma_{ij}=1$)
\begin{align} 
	t_{ij} = \frac{s_{ij}+\eps}{s_{ij}+s_{ji}+2\eps}.
	\label{tij}
\end{align}
We put $\eps>0$ to ensure that $t_{ij}=1/2$ even when $s_{ij}=s_{ji}=0$, e.g. a scoreless tie in a game.

We need to make one more consideration before presenting the final form of the transition matrix in light of the case depicted in Fig.~\ref{figure01}~(d).  Here one could argue that the relationship between the three teams --- $i$ (left) tied with $k$ (center) tied with $j$ (right) --- ought to drive the teams' rankings to be equal due to the transitive property.  We see, however, that the two edges incident upon node $k$ penalizes it by acting as two pathways into it, raising the occupation probability in comparison with the other two teams.  We correct for such a bias via following final form for the Markov transition matrix $\P=\set{P_{ij}\equiv P_{i\from j}}$: 
\begin{align} 
P_{ij}= \alpha_j \cdot \sigma_{ij} \cdot t_{ij} \times k_i^{-1}.
\label{transprob} 
\end{align} 
Here the probability of entering a node $i$ is discounted by its degree $k_i$ (not to be confused with the node labeled $k$ in Fig.~\ref{figure01}~(d)), and $\alpha_j$ is the normalizing factor so that $\sum_{i(\ne j)}P_{ij}=1$.  Finally, the stationary occupation probability vector $\vec{l}$ is the leading eigenvector of $\P$ with eigenvalue $1$, i.e. $\P\vec{l}=\vec{l}$~\cite{pierre1999markov}

For completeness, we note that we could have equally let the walker prefer to move to the stronger node, in which case the occupation probability would represent the node's strength.  This can be achieved by introducing a different transition matrix $\P'=\set{P'_{ij}}$ where we simply set $t'_{ij}\equiv t_{ji}$ from Eq.~\eref{tij}. Then its occupation probability vector which we label $\vec{w}=\set{w_1,w_2,\ldots,w_n}$ satisfies $\P'\vec{w}=\vec{w}$.  Finally, since there is no \emph{a priori} reason to favor one picture over the other, we combine them into $\vec{\ra}=\vec{w}-\vec{l}$ to use as the final strength measure.

Our method has a number of differences from PageRank, Fig.~\ref{figure01}~(b). First, our method allows a bidirectional walk with a gradient defined by the scores, Eq.~\eref{tij}, rendering the random jump component of PageRank unnecessary as long as the network is connected.  Second, the transition probability in our method (Eq.~\eref{transprob}) neutralizes for the degree of the potential target node unlike in PageRank where having a large indegree generally leads to a higher centrality. 

\begin{figure} 
\includegraphics[width=80mm]{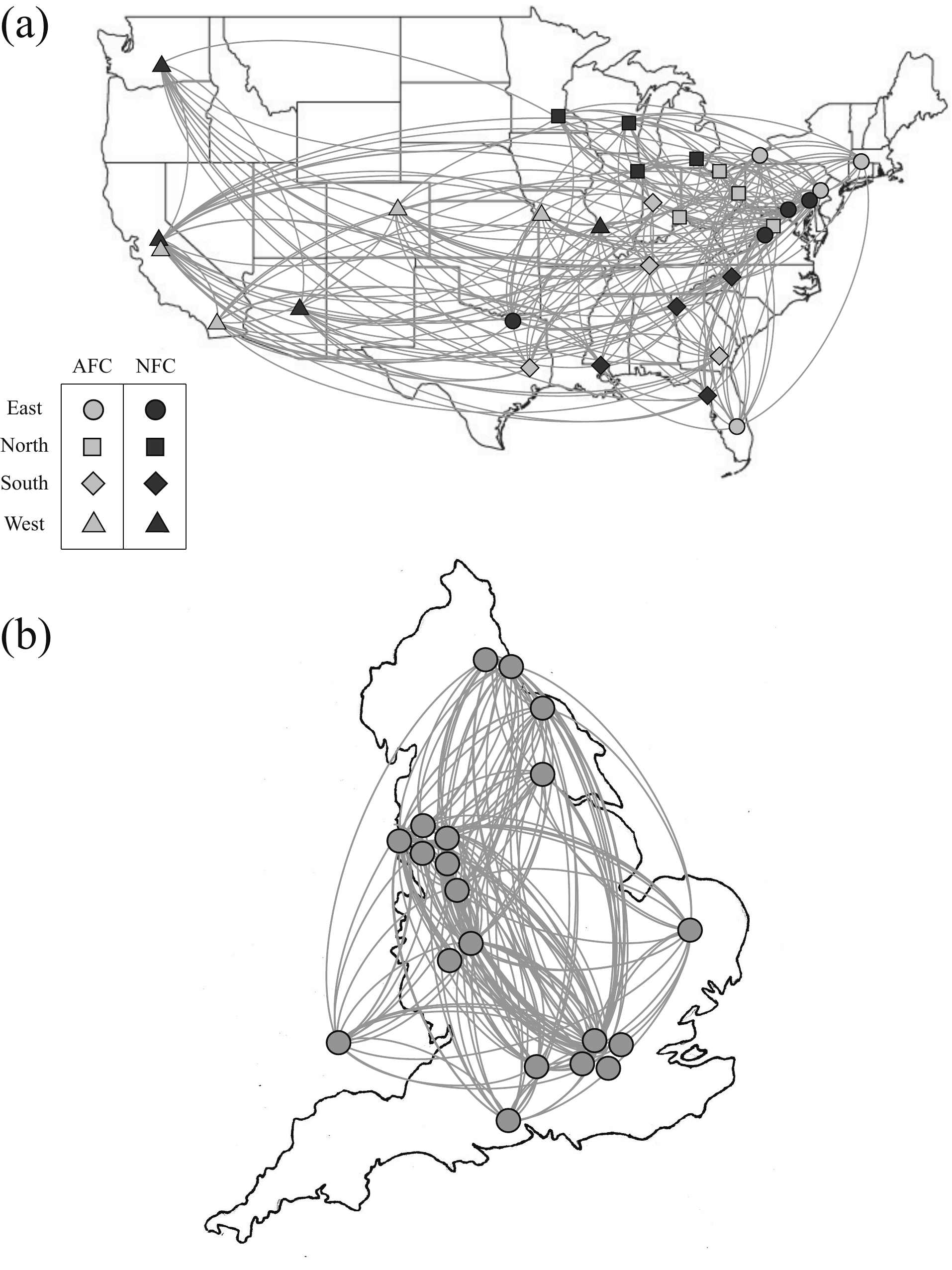} 
\caption{The schedule networks for (a) the National Football League (NFL) in 2013 and (b) the English Premier League (EPL) of 2012--2013. The NFL consists of 32 teams divided equally into American Football Conference and National Football Conference, further divided into four divisions (shaped differently) corresponding to regions in the country. The EPL consists of 20 teams, forming a complete network or a round-robin.} 
\label{figure02} 
\end{figure} 

\section*{Results and Discussion}
To gauge the performance of our method and to better understand its implications we apply it to two real-world competition networks found in sports.  We use two competition networks, the National Football League (NFL, www.nfl.com) of the USA, and the English Premier League (EPL, www.premierleague.com).  The schedules for the NFL (year 2013) and the EPL (year 2012--13) are shown in Fig.~\ref{figure02} as undirected networks.  The NFL consists of 32 teams divided into American Football Conference (AFC) and National Football Conference (NFC) that are further divided into four divisions of 4 teams, respectively.  Annually they play 256 regular-season games (16 games for each team), the outcomes of which act as the the basis of the playoff that culminate in the championship game (called the Super Bowl).  The EPL consists of 20 teams, each playing against each other twice during the season for a total of 38 games for each team.  Such a full network is called ``complete'' or a ``round-robin.''  When two teams play multiple times (in the EPL it happens between every pair of teams, and in the NFL it happens between teams belonging to the same conference) we let $s_{ij}$ and $s_{ji}$ represent the cumulative points, and update the gradients $t_{ij}$ and $t_{ji}$ accordingly.  The necessary condition on $\eps$ in Eq.~\eref{tij} is that it is positive, and we set $\eps=1/2$ here.

\begin{figure*} 
\includegraphics[width=180mm]{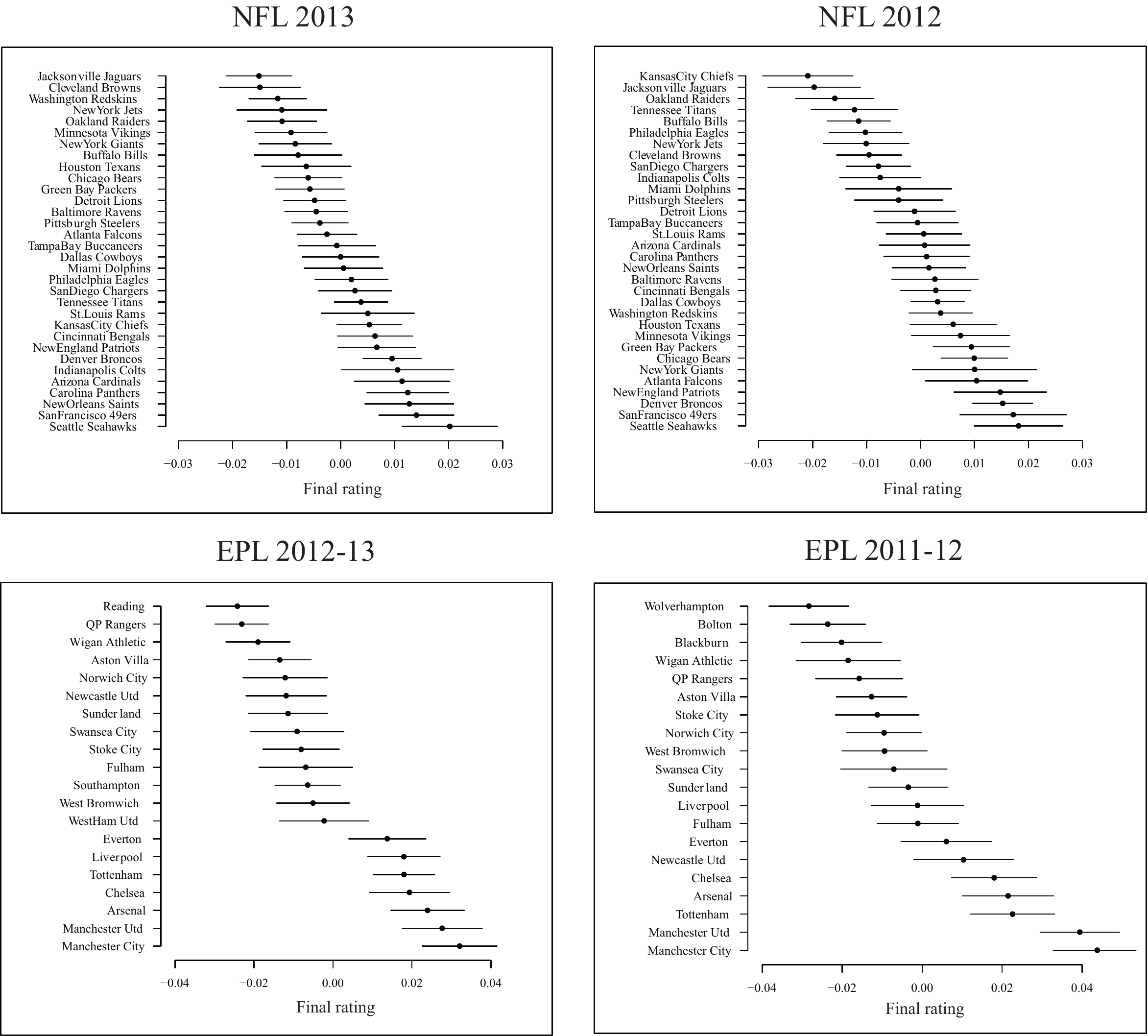} 
\caption{The final regular season ratings $\vec{\pi}$ and rankings of the teams for two professional sports, NFL (top) and EPL (bottom). The errors were estimated using the jackknife method~\cite{efron1979computers,newman1999monte}. In NFL 2013 the top-ranked team (Seattle Seahwaks) enjoyed an exceptionally regular season and won the Super Bowl championship in a dominant fashion.}
\label{figure03} 
\end{figure*}

The final ratings $\rats$ and the rankings of the teams are shown in Fig~\ref{figure03}, with the error bars obtained using the jackknife method~\cite{efron1979computers,newman1999monte}.  We used the 2012 and 2013 regular season data for the NFL, and the 2012--13 and 2011--12 data for the EPL. We first discuss the results from the NFL (top). In 2013 (top left), the Seattle Seahawks show a noticeably high score in comparison with other teams, reflecting the dominance they showed during the regular season; as a matter of fact, they proceeded to defeat every opponent in the postseason and capture the championship. The score of 43--8 against the Denver Broncos in the Super Bowl was the most lopsided in NFL history.  This is not always the case, though, as the error bars suggest. In 2012 (top right) it was the 14th-ranked Baltimore Ravens that won the championship after entering the postseason as the lowest-seeded team. The result was a surprise to many, as their progress to the championship was considered a series of ``upsets'' -- a lower-ranked team defeating a higher-ranked team.  The EPL (bottom) lacks a postseason, but our method agreed with the official EPL ranking system in choosing the four top teams that get to represent the EPL in the UEFA (Union of European Football Associations) Champions League, an annual European competition played between clubs. 

\begin{figure} 
\includegraphics[width=80mm]{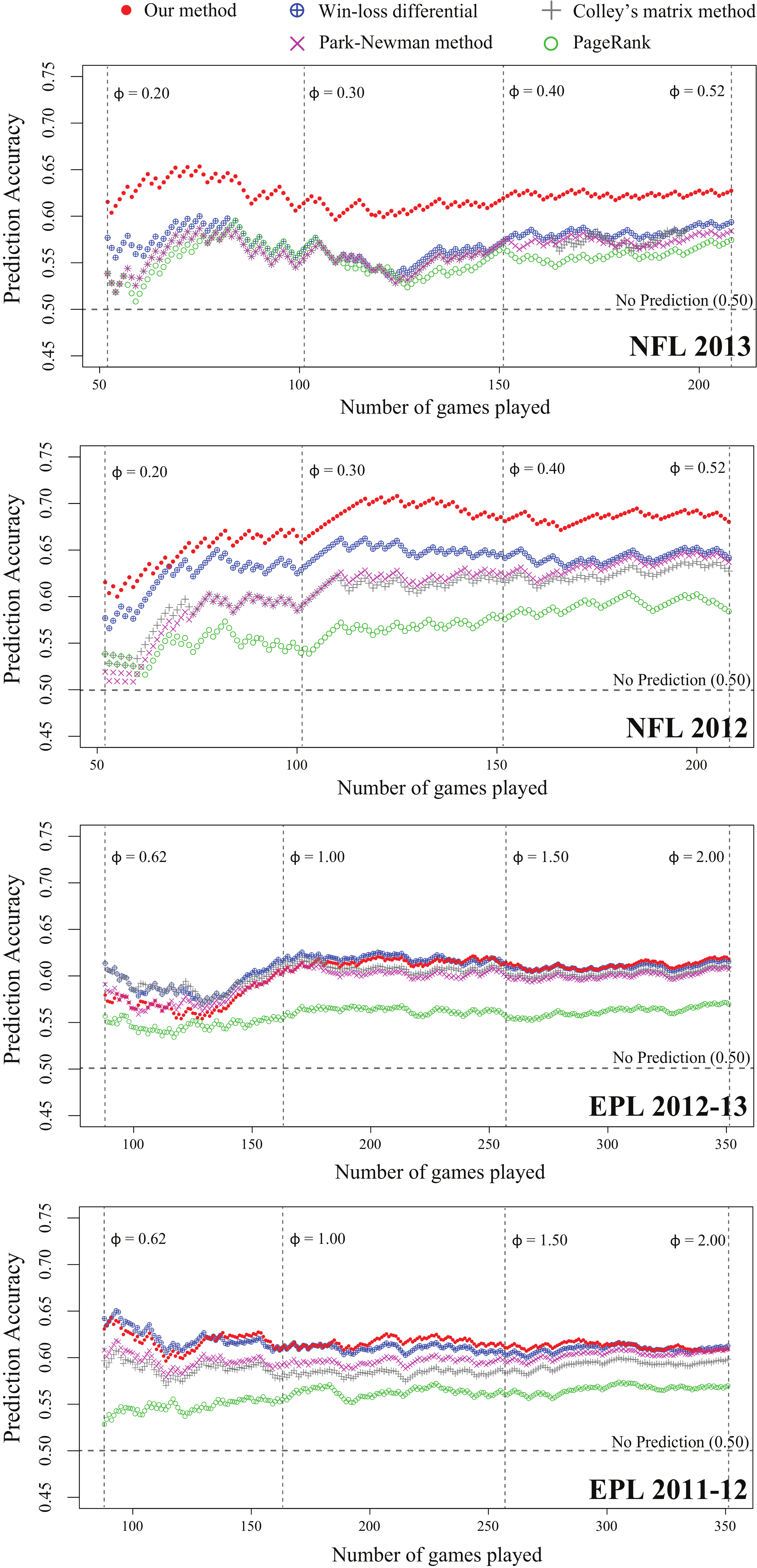} 
\caption{The weekly prediction accuracies of our method and four other methods compared.  Predictions were made based on cumulative data. In the case of NFL (top panels) our method shows a noticeably higher prediction accuracy in comparison with others, while for the EPL (bottom panels) the methods exhibit smaller differences, mainly due to the significantly higher connectance $\phi$.}
\label{figure04} 
\end{figure}

The lack of upsets -- likely noting the stability of the ranking -- in the EPL is likely due to its larger \defn{connectance} $\phi$ defined as
\begin{align}
	\phi \equiv \frac{m}{{\binom{n}{2}}} = \frac{m}{\frac{n(n-1)}{2}},
\end{align}
where $m$ is the number of edges, i.e. the actual games played. Since in the EPL two games are played between every pair of teams, $\phi_{\mathrm{final}}=2.0$, meaning that we have more information on which to base our final predictions.  In general, since our ranking method produces a set of ratings of the teams based on data (i.e. past performance), how well it functions as a predictor of future outcomes is an interesting problem to look at.  We explore this by studying the weekly prediction accuracies  of our method as the season progressed, given by the number of correct wins predicted divided by the total number of games played (a tied game was considered half correct). They are shown in Fig.~\ref{figure04}, given as a function of connectance $\phi$.  For reference, we compare our method with others. As it would be infeasible to perform a comprehensive comparison encompassing all existing methods it is important to select those that are practically impactful or scientifically illustrative. We therefore chose the following four: 
\begin{enumerate} 
\item \textbf{Win--loss differential with tie breaker.} This predicts the team with a higher win--loss margin to win. In case the margins are tied, a ``tie breaker'' is employed by which the team that has scored more net points during the season is predicted to win. This is the official ranking system of the EPL. 
\item \textbf{Park-Newman network ranking method.} Developed by Park and Newman in 2005~\cite{park2005network}, this method ranks teams according to their generalized wins--losses that take into account the number of indirect paths between nodes. They showed that it corresponds to a directional version of Katz centrality~\cite{Katz53a}. 
\item \textbf{Colley's matrix method.} Devised by W.~N. Colley, nodes are ranked by ratings scores calculated from an iterative scheme. This method is notable for being an official computational method to be used in the US college football, and one of the few whose detail is made public~\cite{Colley2002}. 
\item \textbf{PageRank.} In accordance with our definition of the directed edge pointing from the winner to the loser of a game, we interpret the occupation probability as indicating a team's weakness~\cite{brin1998anatomy}. 
\end{enumerate} 

The changes in the weekly prediction accuracies are given in Fig.~\ref{figure04}  for the five methods. We see that our method consistently outperformed others in the NFL, while they were more or less on par in the EPL (except for PageRank that noticeably underperformed): the aggregate prediction accuracies of our method and the four methods (in the order in which they were presented above) were $(0.627,~0.594,~0.584,~0.584,~0.575) $ for NFL in 2013, $(0.680,~0.642,~0.637,~0.627,~0.584)$ for NFL in 2012, $(0.618,~0.615,~0.607,~0.610,~0.570)$ for EPL in 2012--13, and $(0.610,~0.613,~0.610,~0.598,~0.570)$ for EPL in 2011--12. While some methods may perform slightly better than our method in the early stages of the season in the EPL, our method starts to perform equally well or better as the season progresses, reaching a larger $\phi$ value. The comparison results in Fig.~\ref{figure04} appear to indicate the effectiveness of the aspects of our method absent in other methods, namely the score-based gradients for random walk and degree neutralization. We see that PageRank underperforms other method noticeably, which we believe can be attributed to the random jump mechanism working as an indiscriminate equalizer of nodes. 

\begin{figure*} 
\includegraphics[width=180mm]{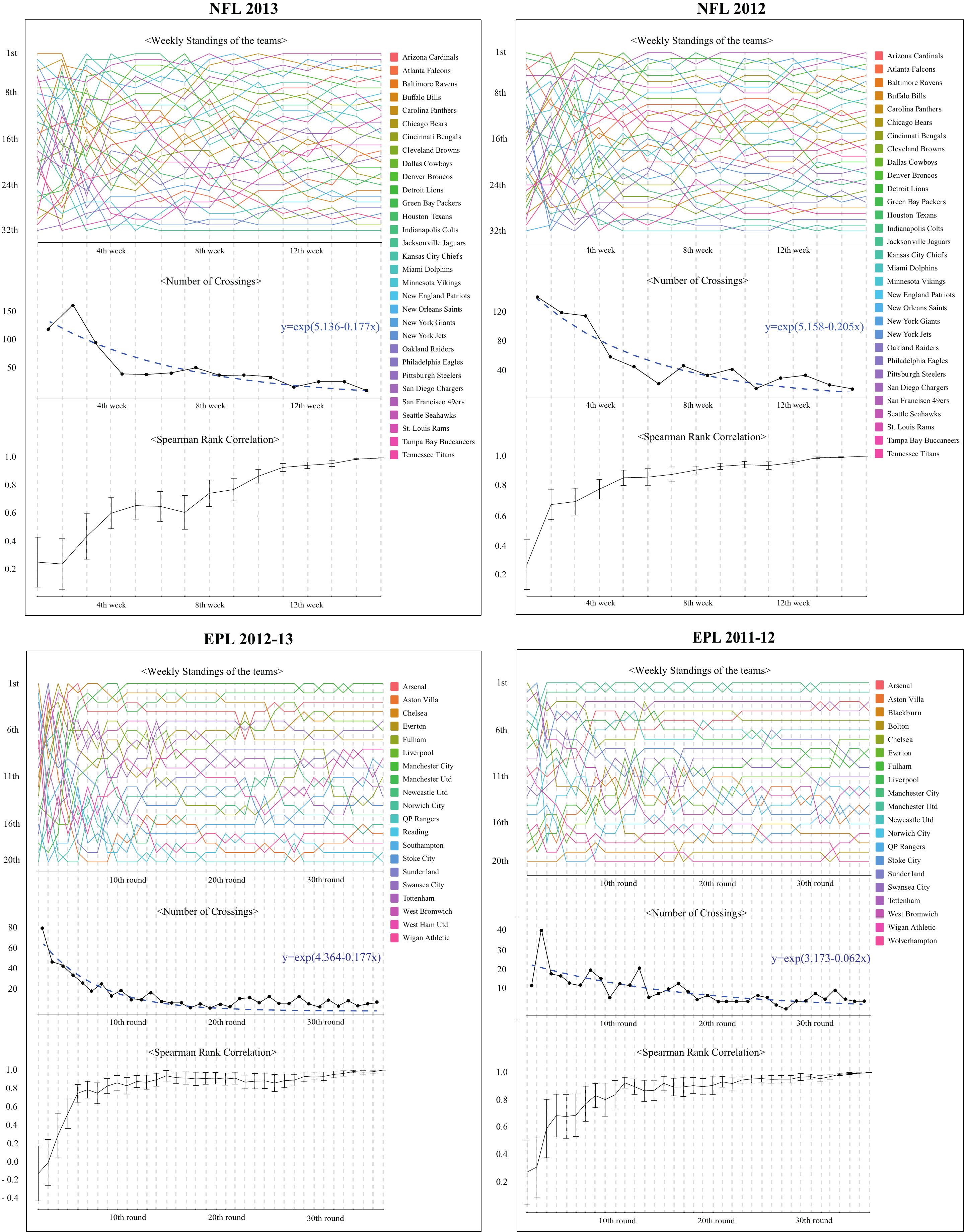} 
\caption{The stabilization of rankings and convergence towards the final ranking. The colored lines (top) track the weekly rankings of teams, which show large fluctuations in the early stages that attenuate as the seasons progress. The fluctuations are quantified by the numbers of line crossings (middle) that show an exponential decrease.  This is also reflected in the Spearman Ranking Correlation of the weekly rankings with the final ones (bottom), which reaches $0.9$ when fewer than $50\%$ of the games have been played with the exception of NFL in 2013, where $64.7\%$ of the games had to be played.} 
\label{figure05} 
\end{figure*} 

We now study how quickly the rankings stabilize and converge towards the final ones as a function of $\phi$.  A ranking of teams produced when $\phi<1$ is often called a partial ranking.  As the seasons progress and games are played more information become available (for us, in the form of $\sij$ and $\sji$), the rankings are likely to stabilize by experiencing fewer and fewer changes.  In Fig.~\ref{figure05} we show the weekly rankings of the teams based on cumulative records, connected by colored lines to show more clearly how the rankings changed over time.  As expected, with the progress of the season the rankings stabilize, evidenced by the decreasing number of line crossings (switching of rankings between teams).  The numbers of line crossings appear to follow an exponential fit $y(x)=\ex{-ax+b}$ for both sports, with a faster rate of decrease in the beginning than near the end.   We can also observe this from the Spearman Rank Correlations (SRC) and the jackknife errors of the weekly rankings with the final ones (so that SRC $=1$ when the seasons end):  We find the points SRC reaches $0.9$, which is the 11th week for NFL 2013 ($\phi=0.387$), 8th week for NFL 2012 ($\phi=0.294$), the 14th week for EPL 2012--13 ($\phi=0.837$), and 11th week for EPL 2011--12 ($\phi=0.679$), each corresponding to only $64.7\%$, $47.1\%$, $35\%$, and $27.5\%$ of the full seasons.  Therefore, the season has to have progressed less than two thirds, often half, to produce a ranking that we can state is substantially similar to the final ones.  This also implies that in the later stages the line crossings occur between teams with closer rankings. We see that it is the middle tier where the line crossings are most common and persistent, showing that they are the most competitive: In the EPL the \#1 and \#2 teams remain stable past midseason (2012–-13) or from the very early stages (2011–-12), and in the NFL we see a similar behavior (albeit slightly weaker) in the top tier.  This also suggests another possible reason for the prediction performances we see in Fig.~\ref{figure04} for the EPL: With connectance $\phi_\mathrm{final}\gtrsim1$ (which is very unusual for real-world networks), enough information is available for even the simplest of schemes, and therefore two ranking methods may not differentiate themselves as readily.  This tells us that our method is more effective on sparser networks, in this case the NFL.  This is in line with the general characteristics of network centralities (some of which are shown in Fig.~\ref{figure01}~(a)) that they become less effective as a network becomes denser and the topology more uniform around each node.

\section{Conclusion}
In this paper, we have introduced a random-walk based ranking system for competition networks.  Our method possessed two properties that render it generally applicable for any network: First, walks were allowed in both directions governed by gradients defined by the edge weights.  Second, the effects of a high degree was neutralized, eliminating unwarranted advantage or disadvantage caused by utilizing the steady-state occupancy as the measure of a node's strength and weakness.  

We applied our method to two popular sports, the National Football League and the English Premier League, to explore the performance and potential uses of our method.  We compared the prediction accuracies of our method with four other methods including the win--loss scheme with a net points-based tie breaker, finding that ours outperforms significantly in the NFL, and is on par in the EPL. We also studied in detail the converging behavior of rankings, finding large early-stage instabilities replaced by smaller-scale fluctuations between mid-range teams.

We found out that the connectance $\phi$ was an important factor in these behaviors, and that our method was more effective when the network was sparser than the EPL network with $\phi<1$.  This does not lessen the necessity of a sophisticated methods such as ours, however: Since most known real-world networks are sparse, the enhanced performance in such cases is an indicator of the value of such methods.

The strength of our method is that it is generally applicable to any system that can be represented as a network of competitions (edges) between components (nodes) where the ranking of nodes is necessary or useful.
We have only explored two networks out of many that can be studied, and we hope to our method applied to more systems in the future, including networks with many-body (not merely two-body, as in our examples) competitions and those from other practical areas of application, such as product recommendation systems based on customer reviews as competitions.

\bibliography{bib_Ranking_Competitors} 
\end{document}